\documentclass[preprints]{article}

\usepackage{graphicx}

\begin{document}

\title{Raychaudhuri Equation,Geometrical Flows and Geometrical Entropy} 
\maketitle

\author[1]{L.P. Horwitz}
\author[2]{V.S. Namboothiri}
\author[3]{G. Varma K}
\author[4]{A. Yahalom}
\author[5]{Y. Strauss}
\author[6]{J. Levitan}\\ \\
\footnote{larry@tauex.tau.ac.il}
\thanks{\textit{School of Physics, Tel Aviv University, 69978 Ramat Aviv, Israel\\ Department of Physics, Bar Ilan University, 52900 Ramat Gan, Israel\\Department of Physics, Ariel University, 40700 Ariel, Israel}}\\.
\footnote{ramharisindhu@gmail.com}
\thanks{\textit{Center for Science in Society, Cochin University of Science and Technology (CUSAT), Kochi, Kerala,India}}\\
\footnote{d20069@students.iitmandi.ac.in}
\thanks{\textit{Department of Physics,Indian Institute of Technology Mandi,Kamand 175005,India}} \\
\footnote{asya@ariel.ac.il}
\thanks{\textit{Department of Engineering, Ariel University, Ariel 40700, Israel.}} \\
\footnote{yossef.strauss@gmail.com}
\thanks{\textit{Department of Mathematics, Ben Gurion University, Beer Sheba 84105, Israel.}}\\ 
\footnote{levitan@ariel.ac.il}
\thanks{\textit{Department of Physics, Ariel University, 40700 Ariel, Israel.}}\\

\thanks{Keywords:Raychaudhuri Equation, Chaos Theory, Kaluza Klein Theory, Kaluza Klein Cosmology, Geometrical Flow, Geometrical Entropy, Riccati equation.}

\hyphenation{preprints}

\maketitle
\begin{abstract}
Raychaudhuri equation is derived by assuming geometric flow in space-time $M$ of $n+1$ dimensions. The equation turns into a harmonic oscillator form under suitable transformations. Thereby a relation between geometrical entropy and mean geodesic deviation is established. This has a connection to chaos theory where the trajectories diverge exponentially. We discuss it's application to cosmology and black holes. Thus we present a connection between chaos theory and general relativity.
\end{abstract}

\section{Introduction}
In general relativity the motion of nearby bits of matter is described by the celebrated Raychaudhuri equation or the Landau-Raychaudhuri equation\cite{Raychaudhuri:1953yv,Ehlers:2006aa}. It shows a general validation that gravitation should be a universally attractive  interaction between any two bits of matter in general relativity and also in Newton's theory of gravity. This equation was formulated by Raychaudhuri and Landau independently in  1954\cite{Kar:2008zz,Dadhich:2005qr}. Later it became a fundamental lemma in proving the famous Hawking-Penrose singularity theorems and in studying exact solutions of Einsteins equations in general relativity\cite{Senovilla:2014gza,Penrose:1964wq}. 

The famous scientist Saurya Das, has proposed in the field of Quantum theory a Raychaudhuri equation where the usual classical trajectories are replaced by Bohmian trajectories. Bohmian trajectories do not converge and thus the issue of geodesic incompleteness, singularities such as big bang or big crunch can be avoided[12,14]. In this paper we treat the classical geometrical flow as a dynamical system in such a way that the Raychaudhuri 
equation becomes the equation of motion and that the action can be used to quantize the dynamical system. Classical chaos is essentially
characterized by the exponential divergence of neighboring trajectories inducing a high degree of instability in the orbits
with respect to initial conditions.

\par The Raychaudhuri equation is the basis for deriving the
singularity theorems. The study is expected to show the effect such a quantization will have on the geometrical flow,and as part of the process it can be shown that a quantum space-time is non-singular.The existence of a conjugate point is a necessary condition for the occurrence of singularities.However it is possible to demonstrate that conjugate points cannot arise because of the quantum effects. An intriguing result obtained was that the Raychaudhuri equation can be written in a harmonic oscillator form under suitable transformations. Here a new quantity called geometrical entropy $S = ln \chi(x)$ can be defined where $\chi(x)$ represents the distance between two nearby geodesics. We have expressed the above
equations in terms of the entropy which by transformation to a Ricatti-type equation becomes similar to
the Jacobi equation. We have recently proved that the geodesic deviation equation of Jacobi becomes unitarily equivalent
to that of a harmonic oscillator. In this way, a connection between general relativity and  chaos theory is established [8-10].
 
The connection can be further investigated by the addition of gauge fields in the metric. Here too the
Raychaudhuri equation and the geodesic equation acquire the harmonic oscillator-form under suitable transformations. However, the convergence and divergence criteria get modified by the effect of the gauge field. In this case the particles deviate from the geodesics. A point to be noted when adding a gauge field into the picture is that the particle no longer follows a geodesic. According to the work by S.G.Rajeev, the Riemannian
geometry is a particular case of Hamiltonian Mechanics. He explores the links between Riemannian
geometry and Hamiltonian Mechanics by changing the form of the Hamiltonian through the addition of a scalar field or
vector field and investigates the corresponding change in the geometry(change in curvature and Ricci
tensor).
 \section{Raychaudhuri Equation from geometric flow}
 We study the congruence of a test particle moving on an n+1 dimensional space-time M. We use the proper time ($\tau$) for this particle  as a dynamical foliation parameter so as to foliate the space-time into topology $T \times R$. Here $T$ is a Riemannian Manifold with a metric $g_{\alpha\beta}$ that projects any vector field into the manifold. We also define $H(T)$, a hyper-surface in the transverse manifold that the world-lines intersect at time $\tau$. The volume of that hyper-surface is given by: 
 \begin{equation}
Vol=\int_{\sigma_{\tau}}\sqrt{det\:g}d^{n}x
\end{equation}
we consider the velocity field of the test particle in the congruence to be normal to the n-dimensional transverse manifold $H(T)$. The gradient of velocity is a second rank tensor having three parts : the symmetric traceless part, the antisymmetric part and the trace. The three parts define the shear, the rotation and the expansion of the flow. We consider the cross-sectional hypersurface $\sigma_{\tau}$
as a dynamical system, and its volume as the dynamical degree of freedom. We define the dynamical degree of freedom as
\begin{equation}
 \rho(\tau) = 2\int\sqrt{det\:g}d^{n}x
\end{equation}
 we define the dynamical evolution of the metric as:
 \begin{equation}
\partial_{\tau}g_{\alpha \beta}= \theta_{\alpha \beta}\\
= 2\sigma_{\alpha \beta}+\frac{2}{n}g_{\alpha \beta}\theta
 \end{equation}

Multiplying both sides by $\sqrt{det\: g}$
and using $\delta(det \:g) = g\:g_{\alpha \beta}\:\delta g_{\alpha \beta}$, we get a very important result
\begin{equation}
 \dot{\rho} = \frac{2}{n}\rho \:\theta  
\end{equation}
Using the Lagrangian $L=  (\frac{n}{4}\,\frac{1}{\rho}\, \dot \rho^2 \, -\rho (\mathcal R\, -(\dot\xi^{\alpha})_{;\alpha})-  V_\sigma(\rho))$, we define the action
\begin{equation}
 S( \rho, \dot \rho) = \int d\tau (\frac{n}{4}\,\frac{1}{\rho}\, \dot \rho^2 \, -\rho (\mathcal R\, -(\dot\xi^{\alpha})_{;\alpha})-  V_\sigma(\rho)) 
 \label{action}
\end{equation}

Where $\mathcal{R}$ is the Raychaudhuri scalar, $\mathcal{R}:= R_{\mu \nu} \xi^\mu\xi^\nu$, and~$V_\sigma(\rho)$ is the shear potential that satisfies the equation

\begin{equation}
\frac{ V_\sigma(\rho)}{\partial \rho} = 2\sigma^2
\end{equation}
We can express canonical conjugate momentum as
\begin{equation}
\Pi = \frac{\delta L}{\delta \dot \rho}  =\frac{n}{2}\, \rho^{-1} \dot \rho, \nonumber \\
= \frac{n}{2}\, \rho^{-1} (\frac{2}{n} \rho\, \theta) = \theta.
\end{equation}
Thus, as one would expect, the expansion parameter is the conjugate momentum to
the dynamical degree of freedom. We proceed further by computing the variation
\begin{equation}
\frac{\delta L}{\delta \rho} =-\frac{n}{4}\rho^{-2} \dot \rho^2 - 2 \sigma^2 - \mathcal R+(\dot\xi^{\alpha}_{;\alpha})= \frac{n}{4} \left[ \rho^{-2} \left(\frac{4}{n^2} \rho^2 \theta^2 \right) \right]- 2 \sigma^2 - \mathcal R+(\dot\xi^{\alpha}_{;\alpha})= -\frac{1}{n}\: \theta^2 - 2 \sigma^2 - \mathcal R +(\dot\xi^{\alpha})_{;\alpha} \end{equation}
 Using the Euler-Lagrange equation, we can rewrite the Raychaudhuri equation as
\begin{equation}
\frac{d \theta}{d \tau} = \frac{\delta L}{\delta \rho}
\end{equation}
\begin{equation}
  \dot \theta = -\frac{1}{n}\, \theta^2 - 2 \sigma^2 - \mathcal R+ (\dot\xi^{\alpha})_{;\alpha}
  \end{equation}
We can define the Hamiltonian as
\begin{equation}
H = \frac{1}{n} \rho\, \theta^2  + (\mathcal R-(\dot\xi^{\alpha})_{;\alpha}) \rho +V_\sigma(\rho)
\end{equation}. Thus the derivation of Raychaudhuri equation from geometric flow without acceleration term can be find.

\section{Raychaudhuri equation in harmonic oscillator form}

Let us consider Raychaudhuri equation without the acceleration term ($\dot\xi^{\alpha}_{\alpha}$ set to zero). 

\begin{equation} \label{ray-simp}
 \frac{d\theta}{d\tau} + \frac {\theta^2}{3}  + \sigma^2    =  -R_{\alpha \beta}\xi^{\alpha}\xi^{\beta}
\end{equation}
In order for the LHS  to be negative it must fulfill  the condition  $\frac{d\theta}{d\tau}< - \frac{1}{\theta^2}$  which finally 
leads to the inequality
\begin{equation}
 \frac{1}{\theta(\tau)} \geq \frac{1}{\theta_0} + \frac{1}{\theta{\tau}}
\end{equation}
\par One  can infer that any initially converging hyper-surface-orthogonal congruence must continue to converge and within a finite proper time $\tau \leq - 3\theta_0^{-1}$ must lead to crossing of the geodesics. Since the Strong Energy Condition(SEC) causes gravitation to be attractive, matter obeying the SEC cannot cause geodesic deviation,on the other-hand it will increase the rate of convergence. Since entropy is defined as the average convergence/divergence of the geodesics in a congruence, the SEC will cause further decrease in entropy. If we set $\theta = 3 \chi\prime / \chi$  the Raychaudhuri equation is transformed to
\begin{equation} \label{ray-ricatti}
 \frac{\partial^2\chi}{\partial\tau^2} + \frac{1}{3}\left( R_{\alpha\beta} \chi^\alpha \chi^\beta + 
 \sigma^2\right) \chi = 0,
\end{equation}
which is a harmonic oscillator equation \cite{Kar:2006ms}.  

As pointed out above, $\theta$
may be identified with the derivative of the entropy, so that
the entropy will be of the form $S= Ln \chi$. Here, $\chi$ may be identified with an effective or average geodesic deviation. 

Recently, Kar and Sengupta have shown that the condition for geodesic convergence is the existence of zeroes in $\chi$ 
at finite 
values of the affine parameter($\tau$), and they argue that convergence occurs if
\begin{equation}
 R_{\alpha\beta} \xi^\alpha \xi^\beta + \sigma^2 - \omega^2 \geq 0
\end{equation}

Most of the physical matter fields satisfy the strong energy conditions which state that for all time like vectors $U$, the inequality holds
\begin{equation}
T_{\mu\nu}U^{\mu}U^{\nu}\geq \frac{1}{2}Tg_{\mu\nu}U^{\mu}U^{\nu}
\end{equation}
It follows, when(SEC) holds the term $R_{\mu \nu}U^{\mu}U^{\nu}$ is always positive. Furthermore, note that the shear and the rotation are spatial vectors and consequently $\sigma_{\mu \nu}\sigma^{\mu \nu} \geq 0$ and $\omega_{\mu \nu}\omega^{\mu\nu} \geq 0$. As mentioned above $\omega_{\mu \nu}=0$ is zero if and only if the congruence is hyper-surface orthogonal. If that is satisfied the Raychaudhuri equation simplifies to the form:
\begin{equation}
\frac{d\theta}{d\tau}+ \frac{1}{3}\tau^{2}+\sigma^2= -R_{\mu\nu}U^{\mu}U^{\nu}
\end{equation}
In order for the left hand side to be negative it must fulfill the condition $\frac{d\theta}{d\tau}<-\frac{1}{3}\theta^2$ which finally leads to the inequality 
\begin{equation}
\frac{1}{\theta(\tau)} \geq \frac{1}{\theta_{0}} + \frac{1}{3}\tau
\end{equation}
If we set $\frac{3F'}{F}$ the Raychaudhuri equation is transformed to 
\begin{equation}
\frac{d^2F}{d\tau^2}+ \frac{1}{3}(R_{\mu \nu}U^{\mu}U^{\nu}+\sigma^2-\omega^2)F=0
\end{equation}
Which is a harmonic oscillator equation. We have recently proved that the geodesic deviation equation of Jacobis unitarily equivalent to that of harmonic oscillator. The expansions rate of growth of the cross-sectional area orthogonal to the bundle of geodesics. Increase/decrease of this area is same as that of divergence/convergence of the geodesics. The average growth of the cross-sectional area is the same as that of the geodesics. The average growth of the cross-sectional area is compatible with the average geodesic deviation. Kar and the Sengaupta \cite{Kar:2006ms} have shown that the condition for geodesic convergence is the existance of zeros in $\ln F$ at finite values of the affine parameter, and they argue that convergence occurs if $R_{\mu \nu}U^{\mu}U^{\nu}+\sigma^2-\omega^2$. Here shear increases convergence and rotation obstructs convergence.

\section{Raychaudhuri equation in harmonic oscillator form: with the acceleration term}
The Raychaudhuri equation in harmonic oscillator form can be written as
\begin{equation} 
  \frac{\partial^2\chi}{\partial\tau^2}  + \frac{1}{3}(\sigma^2 +R_{\alpha \beta}\xi^{\alpha}\xi^{\beta}-(\dot\xi^{\alpha})_{;\alpha})\chi=0
\end{equation}
and convergence occurs if:
 \begin{equation}
\sigma^2 +R_{\alpha \beta}\xi^{\alpha}\xi^{\beta}-(\dot\xi^{\alpha})_{;\alpha} \geq 0
 \end{equation}
This clearly shows that the velocity field has a significant role in the convergence or divergence of world-lines.
\par Let us study this effect in more detail. The acceleration term causes the particle to deviate from geodesic. Therefore it has a logarithmic relation to entropy which increases when geodesics diverge.

\par To give a clear picture, we consider the Kaluza Klein cosmology.
The Kaluza Klein metric is given by, $g_{AB}, A,B=0,1,2,3,5 $ with the  electromagnetic potential
\begin{equation}
g_{AB}=
  \left[ {\begin{array}{cc}
 g_{\alpha\beta}+\alpha^{2}\:g_{55}A_{\alpha}A_{\beta} &\alpha_{0} g_{55}A_{\alpha}\\ 
  \alpha_{0} g_{55}A_{\alpha}  & g_{55} \\
  \end{array} } \right]
\end{equation}
where $g_{\alpha \beta}$ is the 4-dimensional metric and $A_{\alpha}$ is the electromagnetic potential. Now the space-time interval becomes
\begin{equation}
 dS^2=g_{\alpha\beta}\ dx^{\alpha}dx^{\beta}-g_{55}\ (dx^5+ \alpha_{0} A_{\alpha}dx^{\alpha})^2.  
\end{equation}
We also have 
\[ 
g^{AB}=
  \left[ {\begin{array}{cc}
 g^{\alpha \beta} &-\alpha_{0} g^{\alpha\beta}A_{\alpha}  \\
  -\alpha_{0} g^{\alpha\beta}A_{\alpha}  & 1/g_{55}+\alpha_{0}^{2}\:g^{\alpha\beta}A_{\alpha}A_{\beta} \\
  \end{array} } \right].\]
This provides a space-time with electromagnetism and gravity unified. The  geodesic equation in five-dimensional space-time,  
\begin{equation}
\frac{d^2z^A}{dS^2}+{\Gamma}^A_{B C}\frac{dz^B}{dS}\frac{dz^C}{dS}=0  
\end{equation}
 can be transformed by applying cylindrical condition on the metric as
\begin{equation}
\frac{d^2z^{\alpha}}{dS^2}+{\Gamma}^{\alpha}_{\beta\lambda}\ \frac{dz^{\beta}}{dS}\ \frac{dz^{\lambda}}{dS}=a\alpha_{0} F_{\alpha\beta}\frac{dz^{\beta}}{dS}\\
+\frac{1}{2} 
\frac{a^2}{g_{55}^2}g^{\alpha\lambda}\left({\partial}_{\lambda}g_{55}\right), \end{equation}
where
\begin{equation}
a=g_{5\alpha}\frac{dz^\alpha}{dS}+g_{55}\frac{dz^5}{dS}, \nonumber 
\end{equation}
and $a$ is a constant along the $5-D$  world line. In our case $g_{5\alpha}$  is non zero since we have included electromagnetic fields. We assume that
$$ \frac{g_{5\alpha}}{g_{55}}=\alpha_{0} A_{\alpha}(x) $$
where $\alpha_{0}$ is determined by
$$ \alpha_{0}= \frac{q}{amc}.$$
Here, the electromagnetic potential emerges out of $g_{5\alpha}$. Let us now consider Raychaudhuri equation in four dimensions.\
\begin{equation}
\frac{d\theta}{dS}=-\frac{{\theta}^2}{3}-{\sigma}_{\alpha\beta}{\sigma}^{\alpha\beta}+{\omega}_{\alpha\beta}{\omega}^{\alpha\beta}
-R_{\alpha\beta}\xi^{\alpha}\xi^{\beta}+(\dot{\xi}^{\alpha})_{;\beta}
\end{equation}
 where $\displaystyle \xi^{\alpha}=\frac{dz^{\alpha}}{dS}$, $2{\sigma}^2={\sigma}_{\alpha\beta}{\sigma}^{\alpha\beta}$, $2{\omega}^2={\omega}_{\alpha\beta}
{\omega}^{\alpha\beta}$.  The cosmological constant $\Lambda$ is set to zero. 
Since $$\dot{\xi^{\mu}}=\frac{d^2z^{\alpha}}{dS^2}+\Gamma^{\alpha}_{\alpha \beta}u^{\alpha}u^{\beta},$$  the last term in Eq.(7) can be written as 
\begin{equation}
(\dot{\xi}^{\alpha})_{\beta}=-\frac{a^2}{2}\ g^{\alpha\rho} D_{\alpha}({\partial}_{\rho}\frac{1}{g_{55}})+a \alpha_{0} D_{\alpha}(F_{\alpha\beta}\frac{dz^{\alpha}}{dS}) 
\end{equation}
with  $D_{\alpha}$ as the covariant derivative. 
The vorticity $\omega$  and $\sigma$ induces expansion and contraction respectively. It is useful to note that $-D_{\alpha}({\partial}_{\rho}\frac{1}{g_{55}})$ is positive for static spherically symmetric space-time in five dimensions without electromagnetism.  
 The additional term, $-D_{\alpha}({\partial}_{\rho}\frac{1}{g_{55}})$ is positive for static spherically symmetric space-time in five dimensions without electromagnetism. Considering a case with $R_{{\alpha}{\beta}}\xi^{\alpha}\xi^{\beta}>0$ and $\omega=0$
we get 
\begin{equation}
\frac{d\theta}{dS}\leq\frac{1}{4}\theta^2-\frac{a^2}{2}\ g^{\mu\rho}D_{\alpha}({\partial}_{\rho}\frac{1}{g_{55}})+a \alpha_{0} D_{\mu}(F_{\alpha \beta}\frac{dz^{\beta}}{dS})
\end{equation}

For a static spherical symmetric metric,
$-\frac{a^2}{2}\ g^{\mu\rho}D_{\alpha}({\partial}_{\rho}\frac{1}{g_{55}})$ is always positive\cite{Parthasarathy:2017nye}. This indicates that a scalar field will always defocus world lines. Thus in Kaluza Klein cosmology, scalar field always creates a defocus of worldlines and we get a bouncing model of universe\cite{Parthasarathy:2020pmw}.

\section{Conclusion}

The formalism that we have developed can be applied to any physical system where the equation for geometrical flow is valid. This can also be applied to cosmology with a scalar field.
\par The physical significance of geometrical entropy is that, it represents the chaotic behavior of world-lines that are trying to converge or diverge. This can be observed in cosmology where geodesics try to converge near big-bang singularity. But scalar fields try to inhibit the convergence and causes divergence. Thus there is  possibility of bouncing model of the universe in classical theory itself. Further studies are possible in charged black holes.


\begin{thebibliography}{00}


%\cite{Raychaudhuri:1953yv}
\bibitem{Raychaudhuri:1953yv} 
  A.~Raychaudhuri,
  %``Relativistic cosmology. 1.,''
  Phys.\ Rev.\  {\bf 98}, 1123 (1955).
  doi:10.1103/PhysRev.98.1123
  %%CITATION = doi:10.1103/PhysRev.98.1123;%%
  %271 citations counted in INSPIRE as of 26 Mar 2020
  %\cite{Kar:2008zz}
  


%\cite{Ehlers:2006aa}
\bibitem{Ehlers:2006aa} 
J.~Ehlers,
  %``A. K. Raychaudhuri and his equation,''
  Int.\ J.\ Mod.\ Phys.\ D {\bf 15}, 1573 (2006).
  doi:10.1142/S0218271806008966
  %%CITATION = doi:10.1142/S0218271806008966;%%
  %11 citations counted in INSPIRE as of 26 Mar 2020
  
\bibitem{Kar:2008zz} 
  S.~Kar,
  %``An introduction to the Raychaudhuri equations,''
  Resonance J.\ Sci.\ Educ.\  {\bf 13}, 319 (2008).
  doi:10.1007/s12045-008-0013-1
  %%CITATION = doi:10.1007/s12045-008-0013-1;%%
  %7 citations counted in INSPIRE as of 26 Mar 2020
  %\cite{Dadhich:2005qr}

\bibitem{Dadhich:2005qr} 
  N.~Dadhich,
  %``Derivation of the Raychaudhuri equation,''
  gr-qc/0511123.
  %%CITATION = GR-QC/0511123;%%
  %42 citations counted in INSPIRE as of 26 Mar 2020
  















%\cite{Senovilla:2014gza}
\bibitem{Senovilla:2014gza} 
  J.~M.~M.~Senovilla and D.~Garfinkle,
  %``The 1965 Penrose singularity theorem,''
  Class.\ Quant.\ Grav.\  {\bf 32}, no. 12, 124008 (2015)
  doi:10.1088/0264-9381/32/12/124008
  [arXiv:1410.5226 [gr-qc]].
  %%CITATION = doi:10.1088/0264-9381/32/12/124008;%%
  %62 citations counted in INSPIRE as of 26 Mar 2020
  %\cite{Das:2013oda}
  
  
  
  
  
  
  
  
  
  %\cite{Penrose:1964wq}
\bibitem{Penrose:1964wq} 
  R.~Penrose,
  %``Gravitational collapse and space-time singularities,''
  Phys.\ Rev.\ Lett.\  {\bf 14}, 57 (1965).
  doi:10.1103/PhysRevLett.14.57
  %%CITATION = doi:10.1103/PhysRevLett.14.57;%%
  %982 citations counted in INSPIRE as of 26 Mar 2020
  
  
    
  %\cite{Kar:2006ms}
\bibitem{Kar:2006ms} 
  S.~Kar and S.~SenGupta,
  %``The Raychaudhuri equations: A Brief review,''
  Pramana {\bf 69}, 49 (2007)
  doi:10.1007/s12043-007-0110-9
  [gr-qc/0611123].
  %%CITATION = doi:10.1007/s12043-007-0110-9;%%
  %84 citations counted in INSPIRE as of 26 Mar 2020



  
  
  %\cite{Weinstein:2015eqa}
\bibitem{Weinstein:2015eqa} 
  G.~Weinstein, Y.~Strauss, S.~Bondarenko, A.~Yahalom, M.~Lewkowicz, L.~P.~Horwitz and J.~Levitan,
  %``Entropy measures as geometrical tools in the study of cosmology,''
  Entropy {\bf 20}, no. 1, 6 (2017)
  doi:10.3390/e20010006
  [arXiv:1504.07855 [gr-qc]].
  %%CITATION = doi:10.3390/e20010006;%%
  %\cite{Horwitz:2007ap}

\bibitem{Horwitz:2007ap} 
  L.~Horwitz, J.~Levitan, M.~Lewkowicz, M.~Schiffer and Y.~B.~Ben Zion,
  %``On the geometry of Hamiltonian chaos,''
  Phys.\ Rev.\ Lett.\  {\bf 98}, 234301 (2007)
  doi:10.1103/PhysRevLett.98.234301
  [physics/0701212 [physics.class-ph]].
  %%CITATION = doi:10.1103/PhysRevLett.98.234301;%%
  %9 citations counted in INSPIRE as of 26 Mar 2020
  
  %\cite{Strauss:2015hfa}
\bibitem{Strauss:2015hfa} 
  Y.~Strauss, L.~P.~Horwitz, J.~Levitan and A.~Yahalom,
  %``Quantum field theory of classically unstable Hamiltonian dynamics,''
  J.\ Math.\ Phys.\  {\bf 56}, no. 7, 072701 (2015)
  doi:10.1063/1.4918614
  [arXiv:1407.5263 [math-ph]].
  %%CITATION = doi:10.1063/1.4918614;%%
%\cite{Rajeev:2017evp}



\bibitem{Das:2013oda} 
  S.~Das,
  %``Quantum Raychaudhuri equation,''
  Phys.\ Rev.\ D {\bf 89}, no. 8, 084068 (2014)
  doi:10.1103/PhysRevD.89.084068
  [arXiv:1311.6539 [gr-qc]].
  %%CITATION = doi:10.1103/PhysRevD.89.084068;%%
  %45 citations counted in INSPIRE as of 26 Mar 2020
  
 
 %\cite{Ali:2014qla}
\bibitem{Ali:2014qla} 
  A.~F.~Ali and S.~Das,
  %``Cosmology from quantum potential,''
  Phys.\ Lett.\ B {\bf 741}, 276 (2015)
  doi:10.1016/j.physletb.2014.12.057
  [arXiv:1404.3093 [gr-qc]].
 %\cite{Alsaleh:2017ozf}
\bibitem{Alsaleh:2017ozf} 
  S.~Alsaleh, L.~Alasfar, M.~Faizal and A.~F.~Ali,
  
  %``Quantum no-singularity theorem from geometric flows,''
  \bibitem{Alsaleh:2017ozf} 
   S.~Alsaleh, L.~Alasfar, M.~Faizal and A.~F.~Ali,
  Int.\ J.\ Mod.\ Phys.\ A {\bf 33}, no. 10, 1850052 (2018)
  doi:10.1142/S0217751X18500525
  [arXiv:1705.00977 [hep-th]].
  %%CITATION = doi:10.1142/S0217751X18500525;%%
  %1 citations counted in INSPIRE as of 27 Mar 2020
  
  
  %\cite{Rajeev:2017evp}
\bibitem{Rajeev:2017evp}
S.~G.~Rajeev,
%``Curvature in Hamiltonian Mechanics And The Einstein-Maxwell-Dilaton Action,''
J. Math. Phys. \textbf{58}, no.5, 052901 (2017)
doi:10.1063/1.4983665
[arXiv:1701.08026 [math-ph]].

%\cite{Parthasarathy:2017nye}

\bibitem{Parthasarathy:2017nye} 
  R.~Parthasarathy, K.~S.~Viswanathan and A.~DeBenedictis,
  %``Classical defocussing of world lines in higher dimensions,''
  Annals Phys.\  {\bf 398}, 1 (2018)
  doi:10.1016/j.aop.2018.08.017
  [arXiv:1702.05231 [gr-qc]].
  %%CITATION = doi:10.1016/j.aop.2018.08.017;%%
  %2 citations counted in INSPIRE as of 26 Mar 2020
  %\cite{Namboothiri:2019sxa}

  

\bibitem{Parthasarathy:2020pmw} 
  R.~Parthasarathy,
  %``Classical defocussing of world lines -- Cosmological Implications,''
  Annals Phys.\ , 168115
  doi:10.1016/j.aop.2020.168115
  [arXiv:2002.12108 [physics.gen-ph]].
  %%CITATION = doi:10.1016/j.aop.2020.168115;%%
\end{thebibliography}
\end{document}